\documentclass[]{aa}  

\usepackage{orcidlink}
\usepackage[version=4]{mhchem}

\begin{document}

\title{Atmospheric diversity of sub-Neptunes from formation with rock, water, and soot}





\author{Caroline Dorn\inst{1}\fnmsep\thanks{Corresponding author: dornc@ethz.ch}
\and Aaron Werlen\inst{1,2}
\and Sean Jordan\inst{1}
}

\institute{
Institute for Particle Physics and Astrophysics, ETH Zurich, CH-8093 Zurich, Switzerland
\and Department of Earth, Planetary, and Space Sciences, University of California, Los Angeles, CA 90095, USA
}

\abstract{
Recent JWST detections of \ce{CH4} and \ce{CO2} in sub-Neptune atmospheres point to a link between atmospheric composition and the nature of planetary building blocks -- rock, water, or refractory carbon (“soot”) -- yet this connection remains poorly understood. Here we investigate how different formation environments shape the coupled interior and atmospheric compositions of sub-Neptunes. We model planets assembled from varying proportions of rock, water, and soot and compute the global chemical equilibrium and the overlying atmospheric structure. 
We find that planets formed from water-poor material produce atmospheres strongly depleted in carbon-bearing species, with log(\ce{CH4}) and log(\ce{CO2}) below $-4$. In contrast, planets assembled from water-rich building blocks naturally develop methane- and carbon-dioxide-rich atmospheres with elevated metal mass fractions and C/O ratios. The presence of refractory carbon (soot) further enhances methane production and can lead to methane-dominated atmospheres. Comparison with JWST observations suggests that water-rich formation is sufficient to explain K2-18\,b and TOI-270\,d with no soot component required, while TOI-421\,b and GJ3470\,b are consistent with water-poor formation inside the water ice line. The ratio \ce{H2O}/\ce{CH4} combined with the mean molecular weight (MMW) provides a powerful two-dimensional diagnostic linking atmospheric composition to formation environment, with departures from the predicted trends explained by water condensation in temperate atmospheres or fractionated atmospheric mass loss.}


 \maketitle
 \nolinenumbers
\section{Introduction}\label{sec:intro}

Recent JWST detections of \ce{CH4} and \ce{CO2} in sub-Neptune atmospheres point to a link between atmospheric composition and the nature of planetary building blocks — rock, water, or refractory carbon — yet this connection remains poorly understood. Traditional mass–radius (M–R) interpretations invoke mixtures of rock, water, and gaseous envelopes to explain observed densities of sub-Neptunes. However, recent studies point to alternative scenarios: planets enriched in refractory organic carbon, termed soot, and formed between the soot line and the water ice line \citep{bergin_exploring_2014}, may represent a common and previously underappreciated class of low-density worlds \citep{li_soot_2025, lin_carbon-rich_2025}. High-precision atmospheric measurements from JWST now provide an opportunity to test these formation scenarios directly.

\citet{li_soot_2025} argue that bodies forming in carbon-rich regions of protoplanetary disks naturally accrete substantial amounts of soot in addition to rock, analogous to the composition of Solar System comets. Soot, together with water ice, may therefore constitute viable components of the building blocks of sub-Neptunes. Because both soot and water have low intrinsic densities, and because bulk compositions inferred from M–R measurements are highly degenerate \citep{rogers_framework_2010,dorn_bayesian_2017}, these materials are plausible and potentially significant constituents capable of reproducing the observed M–R relationships of sub-Neptunes \citep{lin_carbon-rich_2025}. 

In early planet-forming disks, the main carriers of carbon are organic ices and macromolecular refractory carbonaceous dust. The final carbon budget of a planet may undergo substantial processing and volatile loss \citep{bergin_tracing_2015}. The availability of carbon in the upper atmosphere is strongly influenced by the formation location in the disk, the degree of chemical processing, and the fraction of carbon sequestered into the deep interior. Soot-dominated planets, and variants that also incorporate water ice, termed soot-water worlds, may therefore possess chemically distinct interiors and atmospheres enriched in hydrocarbons such as methane. Such compositions can also promote haze and cloud formation \citep{mahapatra_cloud_2017,lee_mineral_2025}, with important implications for atmospheric observability and habitability. 

In parallel, \citet{lin_carbon-rich_2025} investigate sub-Neptunes with carbon-rich interiors consisting of an iron-silicate core, an intermediate carbon layer, and envelopes spanning metallicities from 1 to 1000 times solar. Their models reproduce the M–R relationships and transmission spectra of key exoplanets, including TOI-270 d and K2-18 b. In their framework, the atmospheres exhibit high metallicities and elevated C/O ratios, reaching up to 100 times solar. A central assumption of this model is that the atmospheric C/O ratio is identical to that of the deeper interior and thus to the planetary bulk. However, when chemically active interiors are considered, the C/O ratios of the bulk, interior, and atmosphere can differ from each other by orders of magnitude \citep{seo_role_2024,werlen_atmospheric_2025} because of volatile sequestration (incl. hydrogen, oxygen, and carbon) in the deep interior \citep{kite_atmosphere-interior_2016,dorn_hidden_2021,schlichting_chemical_2022,luo_interior_2024,bower_diversity_2025}.

Taken together, \citet{li_soot_2025} and \citet{lin_carbon-rich_2025} underscore the importance of carbon as a central component in interpreting sub-Neptune interiors and spectra. A key gap remains: neither study explicitly models interior–atmosphere chemical equilibration to connect the composition of planetary building blocks to observable atmospheric signatures.

\begin{figure*}[t]
    \centering
    \includegraphics[width=.7\textwidth, trim=0cm 0cm 0.cm 0cm, clip]{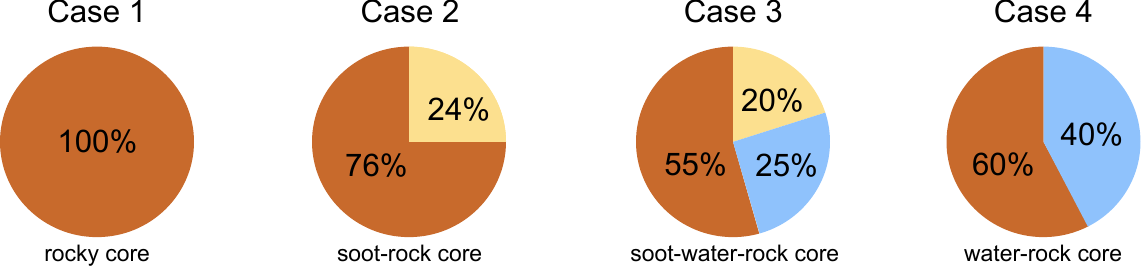}
    \caption{Considered bulk core compositions. Core material excludes priordial H-dominated gas and encompasses all planet building material that is accreted in condensed form. Figure is adapted from \citep{li_soot_2025}.} 
    \label{fig:Cases}
\end{figure*}

This paper aims to fill that gap. We investigate the chemical equilibration mechanisms in soot-enriched planets thereby encompassing chemical reactivity in melts, gases, and between phases, volatile solubility in silicates, partitioning in metals , and outgassing of hydrocarbon species. By coupling thermochemical models with atmospheric structure models, we assess how soot-rich interiors may evolve chemically after formation and how these equilibrium states influence atmospheric signatures. The main underlying assumption is that the interior and the gas are in chemical equilibrium.

We adopt complete interior–atmosphere chemical equilibration as a deliberate end-member. In reality, convection inhibition and mean molecular weight gradients may restrict chemical exchange on a global scale \citep[e.g.,][]{leconte_condensation-inhibited_2017,spaargaren_influence_2020,misener_importance_2022,2022Markham,2024Leconte3d}, such that sub-Neptunes likely achieve equilibration only in certain regions. Current soot-planet models typically adopt the opposite end-member of chemically inactive, decoupled interiors \citep{li_soot_2025,lin_carbon-rich_2025}; our study explicitly examines the consequences of interior–atmosphere equilibration between these two limits. With JWST now delivering transmission spectra precise enough to robustly detect \ce{CH4}, \ce{CO2}, and \ce{H2O} in sub-Neptune atmospheres, these predictions may be directly tested against a rapidly growing observational sample.

This paper is structured as follows. In Section \ref{sec:Method}, we describe the models of both global chemical equilibrium and atmospheric structure. In Section \ref{sec:Results}, we present the predicted trends in atmospheric properties that stem from different bulk core compositions. In Section \ref{sec:Discussion}, we compare model predictions with JWST observations of characterized sub-Neptunes (incl., K2-18\,b, TOI-776\,c, TOI-270\,d, TOI-836\,c, GJ3470\,b, GJ9827\,d, TOI 421\,b, TOI-732\,c, and LP791-18\,c). Finally, we discuss limitations in Section \ref{limit} and conclude in Section \ref{sec:Conclusions}.

\section{Methods}\label{sec:Method}

We adopt the definitions of planet bulk core composition cases as defined in \citet{li_soot_2025}, chemically equilibrate them globally, and investigate how the observable regions of their atmospheres differ between the different bulk compositions. The methodologies for chemical equilibration and atmospheric structure are outlined in Sections \ref{sec:chemical_network} and \ref{sec:atmstruct}, respectively. 

The bulk composition cases that we consider are illustrated in Figure \ref{fig:Cases} and include planetary cores of the following compositions: (1) a purely rocky core, (2) a soot-rock core (24\% soot, 76\% rock), (3) a soot-water-rock core (20\% soot, 25\% water, 55\% rock), and (4) a water-rock core (40\% water, 60\% rock). The four cases represent end-members of planetary cores that were assembled within the soot line (case 1), between soot line and water ice line (case 2), and beyond the water ice line (case 3 and 4). Case 4 ignores any soot and 
can be a proxy for a scenario where interstellar refractory carbon may have been destroyed during heating processes in the protoplanetary disk \citep{anderson_destruction_2017,bergin_tracing_2015}.
Case 4 also represents a reference case to the planet compositions that have been considered in previous formation models outside the water ice line \citep{venturini_nature_2020,burn_water-rich_2024,chakrabarty_where_2024}, especially those that considered chemical equilibration \citep{werlen_sub-neptunes_2025}.

For the rock composition, we follow \citet{young_earth_2023} and use chondritic abundances for C, O, Na, Si, and fix the molar ratios of Mg/Si and Fe/Si to unity \citep[see][for variable refractory element ratios]{grimm_new_2026}.
For soot material, we adapt the composition of C$_{100}$H$_{75-79}$O$_{11-17}$N$_{3-4}$S$_{1-3}$ from \citet{bergin_exoplanet_2023} to a fixed composition of \ce{C100H77O14} for simplicity. Nitrogen and sulfur are not included in the chemical network used here, although their incorporation into global chemical-equilibrium models has been explored in \citet{werlen_role_2026}. A variable soot composition would be more realistic, however our results are only marginally sensitive to the exact soot composition. 

In addition to the planetary core, planets accrete primordial gas early during their formation before the dissipation of the protoplanetary disk. We assume the primordial gas to be composed of 99.9\%~H$_2$ and 0.1\%~CO$_2$ by mole, corresponding to a solar C/O ratio of 0.5 (by moles) \citep{suarez-andres_co_2018}. As the amount of accreted gas can vary mainly depending on planet mass and the temperature environment. We treat the accreted amount of gas as a free parameter and vary the accreted mass fraction from 1-9\%. We show that our results (Section \ref{sec:Results}) are only weakly sensitive in this range. We use planets of 6 and 10 $M_{\oplus}$. For all planets, we assume the temperature at the atmosphere–magma ocean interface (T$_{\text{AMOI}}$) to be equal to 3000 K and the temperature of the silicate-metal equilibration to be 500 K higher. As those temperatures are generally unconstrained, our results of atmospheric mass fraction and composition will, in part, depend on our choice of temperatures. We therefore test all cases also with a temperature of T$_{\text{AMOI}} = 2000$ K and find that although absolute values may vary, the main compositional trends across the investigated core composition cases remain largely unchanged. We highlight differences in the main text whenever relevant.

\subsection{Chemical Thermodynamics}\label{sec:chemical_network}

\begin{figure*}
    \centering
    \includegraphics[width=1\textwidth]{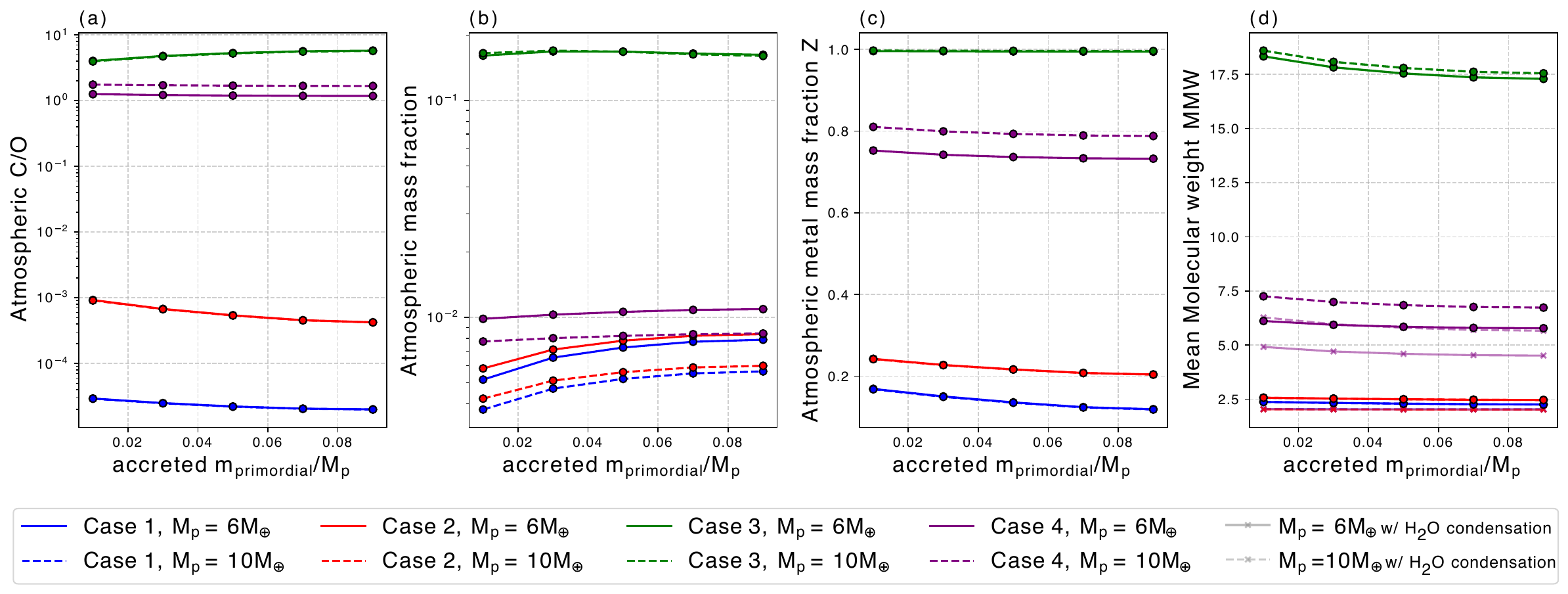}
    \caption{Atmospheric properties of the four modeled planet cases after chemical equilibration: (a) atmospheric C/O ratio, (b) atmospheric mass fraction, (c) atmospheric metal mass fraction $Z$, and (d) mean molecular weight (MMW). The volatile-rich cases (3 and 4) display markedly higher atmospheric C/O ratios and metal mass fractions than the volatile-poor cases (1 and 2). Case 3 also exhibits an atmospheric mass fraction roughly an order of magnitude larger and a significantly higher MMW than in the other cases, including the water-rock case (4). In panel (d), semi-transparent lines with crosses  show the MMW after accounting for water condensation; only Case 4 is visibly affected, as water is negligible in the other cases. }
    \label{fig:bulk}
\end{figure*}

We use the open-source, CPU-based global chemical equilibrium (GCE) model described in \citep{grimm_new_2026}, which has highly improved computational efficiency ($\sim$300x speedup) compared to the original framework developed by \citet{schlichting_chemical_2022}. It also includes a key addition of carbon partitioning in the metal phase \citep{werlen_atmospheric_2025}. The network adopted in this study includes 19 chemical and linearly independent reactions and involves 26 phase components distributed across three coexisting phases: metal, silicate, and gas (see Appendix \ref{sec:appendix_equilibrium}).

We determine the abundances of the 26 phase components, along with the total number of moles in each phase, by simultaneously enforcing chemical equilibrium, mass balance, and elemental conservation in each phase. Our numerical scheme is \citep{grimm_new_2026} has significantly improved in computational efficiency upon the original implementation in \citep{schlichting_chemical_2022}. For a full description of the equilibrium formulation, including the reaction network and governing equations, we refer the reader to the Appendix~\ref{sec:appendix_equilibrium}, as well as \citet{grimm_new_2026}.

Our definition of an astrophysical core, which comprises all material that is accreted in condensed form, reflects the possibility that the two phases of silicates and metals are not necessarily separated by a distinct core–mantle boundary due to considerable sequestration of light elements into the metal. Experimental studies support the solubility of light elements in metallic phases under high-pressure conditions \citep{hirao_compression_2004, terasaki_hydrogen_2009, tagawa_experimental_2021}, a result that is further supported by ab initio calculations \citep{li_earths_2020,luo_interior_2024}. The accumulation of light elements in the metal can lower its density to values comparable to those of silicate, potentially inhibiting gravitational segregation \citep{young_phase_2024, young_differentiation_2025, young_influences_2026}.

\paragraph{\ce{H2} solubility treatment}
We adopt the fixed equilibrium constant ($k_\mathrm{eq}$) prescription for hydrogen solubility in the silicate melt, following \citet{werlen_effects_2026}, in which the dissolved \ce{H2} abundance scales with the fugacity of gaseous hydrogen and therefore increases with pressure. We note, however, that there are alternative formulation using a fixed concentration ratio (k$_\mathrm{D}$) prescription of \citet{schlichting_chemical_2022}, in which the ratio of dissolved to gaseous \ce{H2} remains constant. As demonstrated in \citet{grimm_new_2026}, these two limiting cases can produce substantially different atmospheric mass fractions: the fixed $k_\mathrm{eq}$ prescription allows for more efficient dissolution of \ce{H2} into the silicate melt at high pressure than the fixed-k$_\mathrm{D}$ prescription. Given recent experimental evidence suggesting that hydrogen solubility in silicate melts can reach wt\% levels \citep{miozzi_experiments_2025, horn_building_2025, gilmore_coreenvelope_2026, young_influences_2026}, we have chosen the fixed ($k_\mathrm{eq}$) solubility prescription, and note that this choice represents a key source of model uncertainty.

\begin{figure*}
    \centering
    \includegraphics[width=.9\textwidth]{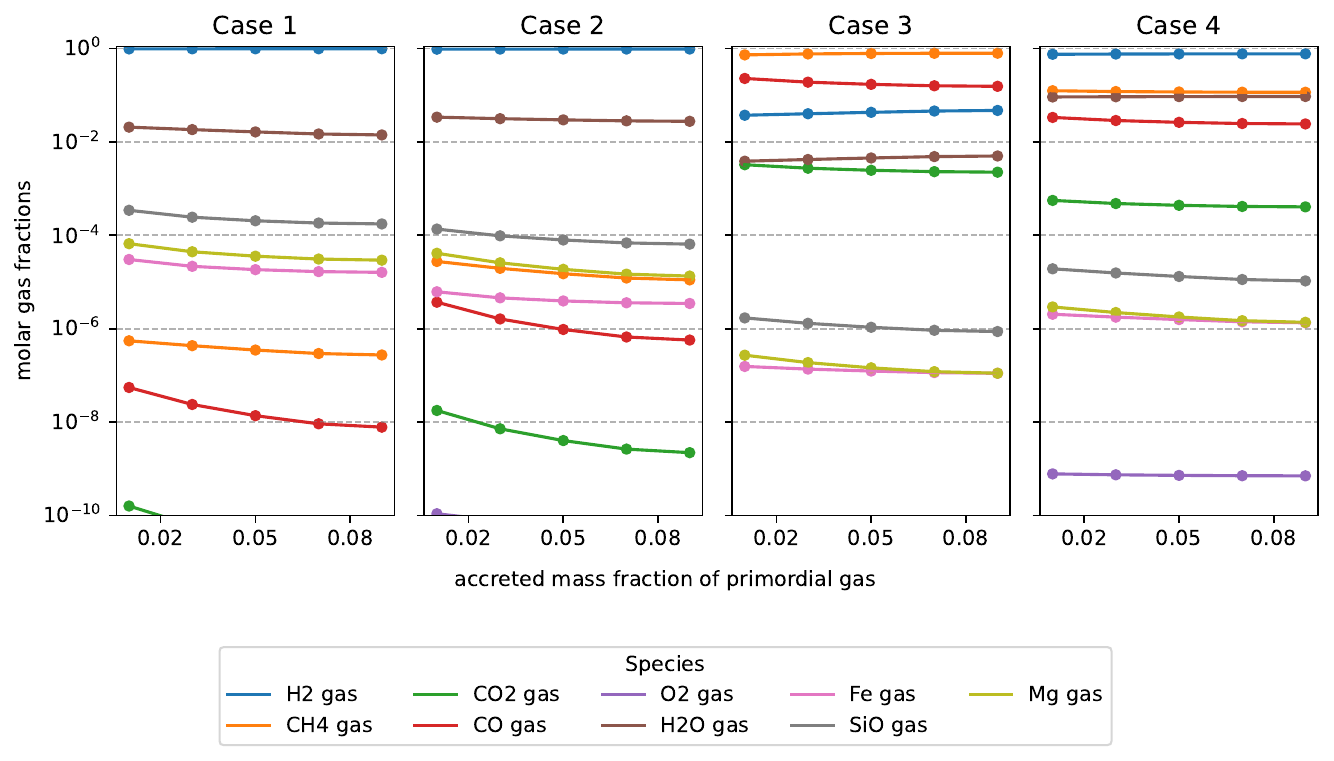}
    \caption{Molar mixing ratios of atmospheric species in equilibrium with the underlying magma ocean. Although the gas composition depends only weakly (to second order) on the fraction of accreted primordial gas, it varies substantially among the four cases. In cases 1 and 2, the atmosphere is dominated by \ce{H2} and \ce{H2O}, whereas case 3 is characterized by \ce{CH4}, \ce{CO}, and \ce{H2}, and case 4 by \ce{H2}, \ce{CH4}, and \ce{H2O}.}
    \label{fig:molarfractions}
\end{figure*}

\subsection{Atmospheric profiles}\label{sec:atmstruct}

We formally couple the deep atmosphere to the observable upper layers of the atmosphere. The deep atmosphere composition and pressure are set by its chemical equilibrium with the magma ocean. Those atmospheric constraints are given as input to a suite of open source atmospheric models: \texttt{FastChem} \citep{stock_fastchem_2018}, \texttt{HELIOS} \citep{malik_helios_2017, malik_self-luminous_2019}, \texttt{VULCAN} \citep{tsai_vulcan_2017}, and \texttt{HELIOS-K} \citep{grimm_helios-k_2021}. These models together enable the simulation of chemical equilibrium, photochemistry, and radiative transfer with species-dependent opacities over the pressure range of the atmosphere and up to pressures of $10^{-8}$ bars.

\texttt{FastChem} computes gas-phase chemical equilibrium under local thermodynamic conditions. \texttt{HELIOS} is a one-dimensional radiative-convective model used to compute pressure-temperature (P–T) profiles. \texttt{HELIOS-K} calculates wavelength-dependent opacities for given gas mixtures. \texttt{VULCAN} is a solver for kinetic reaction networks that incorporates thermochemistry, photochemistry, vertical mixing, and condensation. Together, these models allow us to simulate atmospheric structures from the magma ocean interface to the upper observable atmosphere.

We follow the workflow previously described in \citet{werlen_atmospheric_2025}: The calculated atmospheric compositions output from the GCE model are input to \texttt{FastChem} for the initial P-T conditions representing the atmospheric conditions just above the magma ocean. This output is passed to \texttt{HELIOS-K} to compute gas opacities, which in turn serve as inputs for \texttt{HELIOS} to derive an atmospheric P–T profile. The P-T profile, along with the gas composition above the magma ocean, comprising the lower boundary condition, is input to \texttt{VULCAN} to compute the steady-state chemical structure. We use the output from \texttt{VULCAN} to  recompute the opacities with \texttt{HELIOS-K} and update the P–T structure with \texttt{HELIOS}. This iterative cycle between \texttt{HELIOS}/\texttt{HELIOS-K} and \texttt{VULCAN} is repeated until convergence is achieved in both the gas molar mixing ratios and the P-T profile. In \texttt{VULCAN}, we assume perfect mixing in the deep convective layer and begin calculating chemical kinetics from
$10^{3}$~bar to high altitudes ($10^{-5}$~bar). In our simulations, we adopt a Sun-like stellar spectrum. For individual observed exoplanets, however, the host star’s measured spectrum (if available) is recommended, as assuming a solar analogue can underestimate the incident XUV flux and consequently the formation of photochemical products.

\section{Results} \label{sec:Results}

We compute the chemical equilibrium state for the four core composition cases and investigate how the resulting atmospheric properties depend on formation environment. Figure \ref{fig:bulk} summarizes the bulk atmospheric properties (i.e., C/O ratio, atmospheric mass fraction, metal mass fraction, and mean molecular weight) as a function of the accreted primordial gas fraction.

Planetary cores assembled from volatile-rich material (cases 3 and 4) are characterized by markedly higher atmospheric C/O ratios and metal mass fractions than their volatile-poor counterparts (cases 1 and 2). The C/O ratios of the pure rock (case 1) and soot–rock (case 2) cases fall at least three orders of magnitude below those of the water-rich cases, which reach solar to super-solar values. 
This result may appear counterintuitive, since the addition of water introduces oxygen and might therefore be expected to lower the atmospheric C/O ratio. However, in the water-rich cases the added oxygen is preferentially incorporated into the deep interior. In addition to the dominance of oxygen in silicates, large amounts of oxygen partition into the iron-dominated metal phase, reaching tens of weight percent, while simultaneously yielding increased abundances of \ce{CO} and \ce{CO2} in both the silicate melts and the atmosphere (Figure \ref{fig:molarfractions}).

The dominant atmospheric effect of water addition, however, is a dramatic increase in \ce{CH4} by at least four orders of magnitude (Figure \ref{fig:molarfractions}). The substantial hydrogen supplied by water promotes efficient methane formation \citep[see also][]{steinmeyer_coupled_2026}. As a consequence, carbon is preferentially stored in \ce{CH4}, while much of the added oxygen is sequestered in the interior. The net atmospheric effect of adding water is therefore an increase in C/O, despite the higher total oxygen budget. For a lower temperature  T$_{\text{AMOI}} = 2000$ K than the fiducial T$_{\text{AMOI}} = 3000$ K, a remarkable difference is that the water-rock case 4 shows a C/O ratio an order of magnitude lower, reaching sub-solar C/O values. The C/O ratios of the soot-water-rock case 3 remain high (Figure \ref{fig:bulk}).

The strong increase in \ce{CH4} in the water-rich cases 3 and 4 also explains their higher atmospheric metal mass fractions, which exceed 0.7 and approach unity for the soot–water–rock core (case 3). In contrast, the water–rock case (case 4) remains distinctly below unity and only slightly exceeds values of 0.8.
Planets formed within the water ice line (cases 1 and 2) instead retain hydrogen-dominated envelopes, with water as the second most abundant species, resulting in atmospheric metal mass fractions below 0.3. In all cases, the atmospheric metal mass fractions are only marginally sensitive to variations in T$_{\text{AMOI}}$.

We further find pronounced differences in atmospheric mass fractions when comparing the soot–water–rock case (case 3) with the other cases. The differences are approximately one order of magnitude. Case 3 exhibits characteristically high atmospheric mass fractions of order 10\%, whereas the less volatile-rich cases yield gas mass fractions below 1\%.
For cooler interiors with T$_{\text{AMOI}} = 2000$ K these contrasts become less extreme. The differences decrease to roughly a factor of three, as cases 1, 2, and 4 reach atmospheric mass fractions between 1\% and 5\%.
The atmospheric mass fraction may therefore help distinguish between the two volatile-rich cases (3 and 4), and thereby constrain the role of soot in sub-Neptune interiors, though its discriminating power is temperature-dependent. More robustly, the mean molecular weight (MMW) of the atmosphere differs markedly between cases 3 and 4: case 3 yields MMW  $> 17$ reflecting its methane-dominated composition, while case 4 yields MMW $<7.5$. 

Figure \ref{fig:molarfractions} shows the molar mixing ratios of atmospheric species in equilibrium with the magma ocean across all four cases. Cases 1 and 2 share the same dominant species \ce{H2}-\ce{H2O}-\ce{SiO} in decreasing order. The main differences between case 1 and case 2 are within the minor species: in the soot-rock case, the \ce{CH4} and \ce{CO} mixing ratios are roughly two orders of magnitude higher than in the rock-only case 1, increasing from $\sim 10^{-7}$ to $\sim 10^{-5}$. Despite this relative enhancement, both cases 1 and 2 remain strongly depleted in carbon-bearing species overall. The soot–water–rock case (case 3) differs fundamentally: its atmosphere is methane-dominated at the magma ocean interface, with major species in decreasing order of \ce{CH4}, \ce{CO}, \ce{H2}, and \ce{H2O}. This distinct composition arises from case 3's exceptionally high combined mole fraction of C, O, and H ($\sim$90\%), distributed as approximately 70\% H, 20\% O, and 10\% C. The water-rock case (case 4) has a similarly high combined C-O-H budget, but is carbon-poor in its elemental distribution ($\sim$74\% H, 26\% O, $<$1\% C), producing an \ce{H2}-dominated atmosphere with \ce{CH4}, \ce{H2O}, and \ce{CO} as the next most abundant species.

Taken together, these results establish a clear compositional dichotomy: planets assembled from soot and water are predicted to be methane-dominated worlds, fundamentally distinct from those formed from either component alone, which retain hydrogen-dominated atmospheres. The gas composition across all cases depends only weakly — to second order — on the accreted primordial gas fraction in the range explored (1–9\%), while varying substantially between formation cases. This robustness supports the interpretation of our results as reflecting primarily the bulk composition of the planetary building blocks rather than uncertainties in gas accretion history.

\begin{figure*}
    \centering
    \includegraphics[width=1 \textwidth]{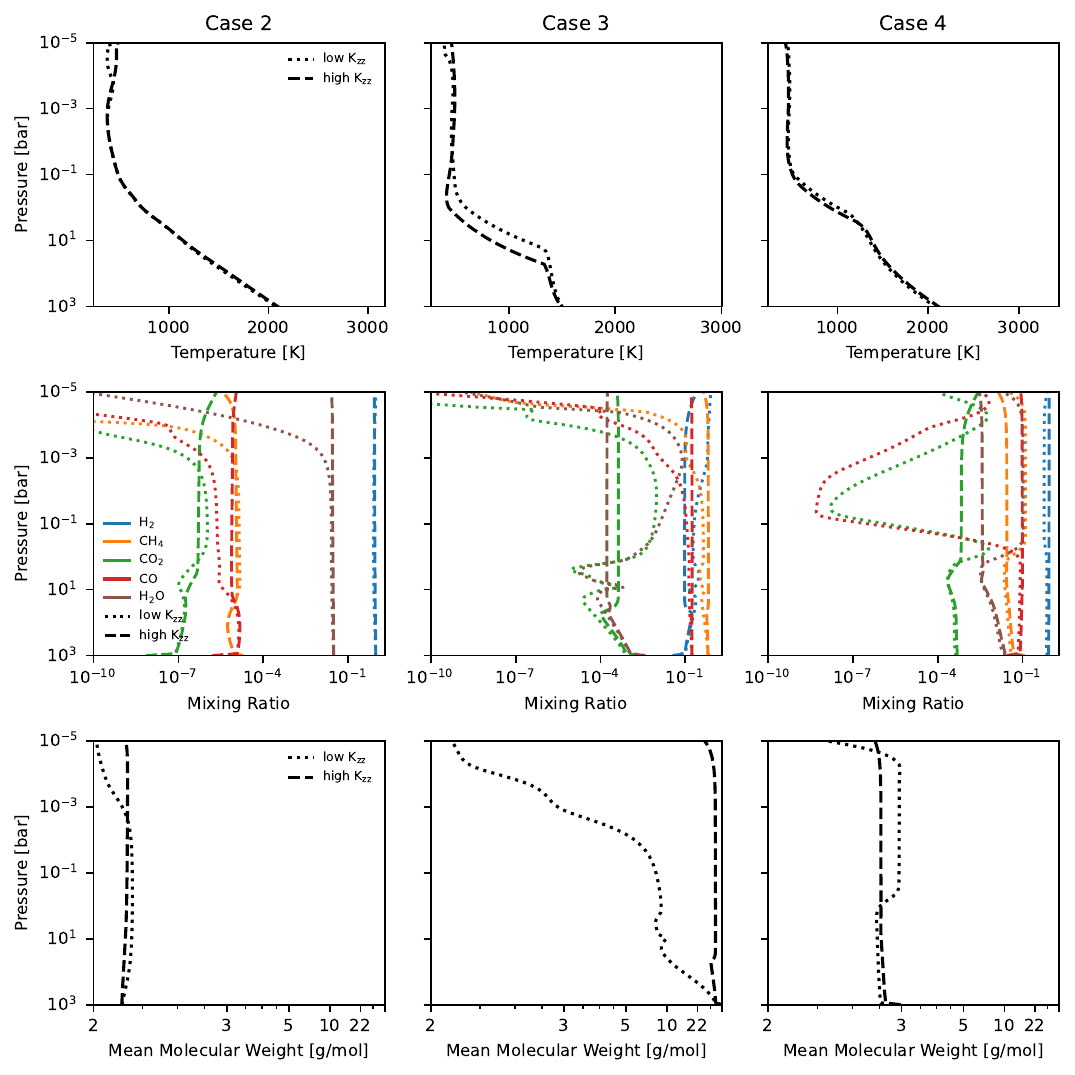}
    \caption{Atmospheric profiles connecting the AMOI interface with the upper 
    observable part of the atmosphere. Profiles are shown for the soot-rock 
    case 2 (left), the soot-water-rock case 3 (middle), and the water-rock case 4 (right). Top: Pressure–temperature profiles before and after convergence. Middle: Vertical mixing ratio profiles of the considered gas species. Bottom: Vertical profiles of MMW. Here, we consider planets of $M_p$ = 6 $M_\oplus$. All panels show results for two different eddy diffusion coefficients, k$_\mathrm{zz} = 10^4$ and $10^7$~cm$^2$s$^{-1}$, which control vertical mixing in the atmosphere. The stellar spectrum is modeled as a blackbody.}
    \label{fig:profiles}
\end{figure*}

\subsection{Upper atmosphere signatures}

Having established how bulk core composition shapes the chemical equilibrium state at the magma ocean interface, we now trace how these deep atmospheric signatures propagate into the upper, observable atmosphere. Figure \ref{fig:profiles} shows the full atmospheric profiles — pressure–temperature structure, vertical mixing ratios, and mean molecular weight — for the soot-bearing cases (cases 2 and 3) and the water–rock case (case 4). We restrict attention to these three scenarios as they span the most chemically distinct outcomes identified in Section 3, and display only the dominant carbon- and oxygen-bearing species. Refractory gases such as \ce{SiO2}, \ce{Mg}, and \ce{Fe} are omitted as they are likely to condense before reaching the observable atmosphere.

The pressure–temperature profiles (top panel) are broadly similar across cases, with one notable exception: below $\sim$0.01 bar, the soot–water–rock case (case 3) develops two distinct convective regions, whereas the soot–rock case (case 2) exhibits only a single deep convective zone. We attribute this structural difference to the higher mean molecular weight and methane-dominated composition of case 3, which alters the thermal gradient in the upper atmosphere. Because refractory condensation is not included in our atmospheric model, however, this feature should be treated with caution and may be modified in more complete treatments.

The middle panel shows the vertical mixing ratios of the dominant carbon- and oxygen-bearing species as a function of pressure. Under weak vertical mixing, the mixing ratios decrease substantially above $\sim 10^{-3}$~bar as molecules become photochemically dissociated. In the strong-mixing regime (high $K_{zz}$), vertical transport continuously replenishes these species from depth, limiting compositional gradients to within roughly one order of magnitude and confining them primarily to the deep atmosphere between 10 and 1000 bar. 

These dynamics are reflected directly in the mean molecular weight profiles (bottom panel). Under strong mixing, the MMW remains nearly constant with pressure across all cases. Under weak mixing, the MMW can vary by up to one order of magnitude in case 3, while changes are moderate for other cases. The sensitivity of the MMW to mixing strength is therefore most pronounced in case 3, where photochemistry dramatically lightens the upper atmosphere.

Despite this sensitivity to vertical mixing, the fundamental chemical differences between cases identified at depth remain discernible within the pressure range probed by transmission spectroscopy ($\sim 10^{-2}$ to $10^{-5}$~bar) \citep[e.g.,][]{benneke_jwst_2024}. Soot–rock planets (case 2) are predicted to exhibit \ce{H2}–\ce{H2O}–\ce{CH4} dominated upper atmospheres in that order, whereas the soot–water–rock case (case 3) may maintain methane dominance with \ce{CH4}–\ce{CO}–\ce{H2} (strong mixing) or \ce{H2}–\ce{CH4}–\ce{H2O} under weak mixing, in both instances retaining methane as a prominent species. Water-rock planets (case 4) similarly transition from \ce{H2}–\ce{CO}–\ce{CH4} under strong mixing to \ce{H2}–\ce{CH4}–\ce{H2O} when mixing is weak. The primary ambiguity introduced by the generally unconstrained strength of vertical mixing is a degeneracy between cases 3 and 4. At low mixing strength, both cases tend toward \ce{H2}–\ce{CH4}–\ce{H2O} compositions. Despite this caveat, the four considered cases occupy sufficiently distinct regions of chemical space that their bulk atmospheric signatures, in particular the \ce{H2O}/\ce{CH4} ratio and the MMW,  remain meaningful diagnostics of formation environment. We now test how well this classification describes the population of sub-Neptunes recently characterized by JWST. In the following, we compare the gas compositions predicted at the AMOI directly with observed sub-Neptune atmospheres, setting aside the additional complexity introduced by atmospheric structure and vertical mixing explored in the present Section.

\subsection{Comparison to characterized sub-Neptune atmospheres}
\label{sec:Discussion}

\begin{figure*}
    \centering
    \includegraphics[width=1\textwidth]{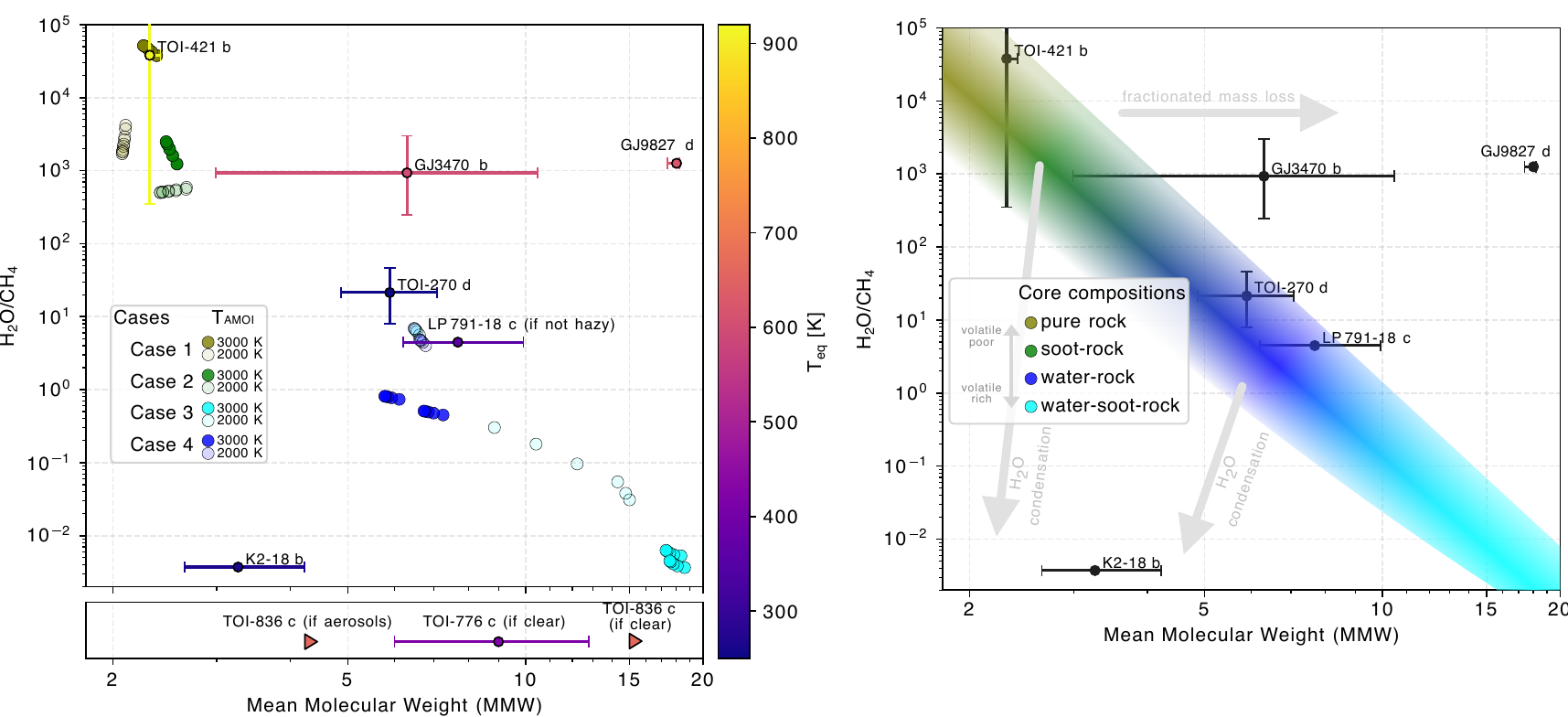}
    \caption{Molar mixing ratios of \ce{CH4}/\ce{H2O} versus mean molecular weight (MMW) as a diagnostic of sub-Neptune formation location - together with inferred atmospheric characteristics of sub-Neptunes from JWST data. Right: Model predictions for all four core composition cases across variations in $T_{\mathrm{AMOI}}$, showing how volatile-poor (cases 1 and 2) and volatile-rich (cases 3 and 4) formation scenarios occupy distinct regions of this parameter space.
Left: Simplified representation of the right panel showing approximate regions occupied by each case, together with locations of characterized sub-Neptune atmospheres. Gray arrows illustrate how two evolutionary processes, i.e., water condensation in temperate atmospheres and fractionated atmospheric mass loss, can displace planets from the main equilibrium trend. TOI-732\,c is omitted as constraints on MMW and water abundances are too weak.}
    \label{fig:overview}
\end{figure*}

Atmospheric characterization of the sub-Neptune population with JWST has revealed a variety of transmission spectra that are either featureless, due to high altitude aerosols or high mean molecular weight atmospheres, or spectra where chemical absorption features can be robustly detected. For sub-Neptunes with chemical features robustly detected from their transmission spectra ($>$\,3\,$\sigma$), the dominant atmospheric absorbers across JWST's NIRISS and NIRSpec wavelength ranges have been either \ce{H2O} or \ce{CH4}, and the atmospheric mean molecular weights have been found to range between $\sim$\,2.3\,g\,mol$^{-1}$ \citep[e.g., TOI-421\,b][]{2025DavenportTOI} up to $\sim$\,18\,g\,mol$^{-1}$ \citep[e.g., GJ9827\,d][]{piaulet-ghorayeb_jwstniriss_2024}. In the classification we present here, \ce{H2O} and \ce{CH4} have also emerged as the dominant atmospheric gases across the four formation cases (in a \ce{H2} background for cases 1, 2, and 4), and the atmospheric mean molecular weights range between $\sim$\,2\,--\,20\,g\,mol$^{-1}$ (Figure \ref{fig:profiles}). It is therefore instructive to see how far our classification can accurately describe the observed population of sub-Neptune atmospheres based on retrieved abundances of \ce{H2O} and \ce{CH4}, and the atmospheric mean molecular weight, which we present in Figure \ref{fig:overview}. For this population analysis we ignore the other carbon bearing species \ce{CO} and \ce{CO2} as the detections of these molecules in sub-Neptune atmospheres, and their reported abundances, have been demonstrated to be model-dependent between different retrieval pipelines \citep[e.g.,][]{schmidt_comprehensive_2025} preventing a trustworthy population-level inference. In contrast, either \ce{CH4}, or \ce{H2O}, or both \ce{CH4} and \ce{H2O} together, have been observed at $>$\,3\,$\sigma$ in every sub-Neptune with any reported chemical detection to date, making these species suitable for population-level chemical analysis.

Across the entire population of sub-Neptunes with robust chemical detections from JWST so far, we find good agreement with our classification system within approximately one order of magnitude. The sub-Neptunes with robust chemical detections, in order of decreasing equilibrium temperature, are: TOI-421\,b \citep{2025DavenportTOI}, GJ9827\,d \citep{piaulet-ghorayeb_jwstniriss_2024}, GJ3470\,b \citep{beatty_sulfur_2024}, TOI-270\,d \citep{benneke_jwst_2024, holmberg_possible_2024, felix_competing_2025}, TOI-732\,c \cite{rigby_jwst_2025}, LP791-18\,c \citep{roy_diversity_2025}, and K2-18\,b \citep{madhusudhan_carbon-bearing_2023, hu_water-rich_2025}. Of this population, TOI-421\,b and GJ3470\,d have low mean molecular weight atmospheres and robust detections of \ce{H2O}. TOI-421\,b has a non-detection of \ce{CH4} and GJ3470\,d has a detection of \ce{CH4} three orders of magnitude less than \ce{H2O} \citep{2025DavenportTOI, beatty_sulfur_2024}. \ce{H2O} in a \ce{H2} background atmosphere with very low \ce{CH4} indicates either case 1 or case 2, therefore TOI-421\,b and GJ3470\,b can be explained by a formation scenario inside the water ice line. 

K2-18\,b and TOI-732\,c also have relatively low mean molecular weight atmospheres, however there are robust detections of \ce{CH4} between $\sim$\,1\,--\,10\% volume mixing ratio, and non-detections of \ce{H2O} \citep{madhusudhan_carbon-bearing_2023, hu_water-rich_2025}. This is consistent with case 4 under the additional consideration that \ce{H2O} can condense out of the observable region of the atmosphere, leading to its non-detection and lowering the mean molecular weight. TOI-270\,d and LP791-18\,c are also both consistent with case 4, with robust detections of \ce{CH4} ranging between $\sim$\,1\,--\,10\% but with intermediate mean molecular weight atmospheres between $\sim$\,5\,--\,8\,g\,mol$^{-1}$ \citep{benneke_jwst_2024, roy_diversity_2025}, which would result from temperatures that are too high for \ce{H2O} to have condensed out. For TOI-270\,d, there may be a similar or greater abundance of \ce{H2O} detected as \ce{CH4} \citep{benneke_jwst_2024}, however the detection significance is below 3\,$\sigma$ and the abundance constraints vary between different retrieval analyses \citep{holmberg_possible_2024, felix_competing_2025}. For LP791-18\,c there is no clear detection of \ce{H2O} although its reported upper limit is high and not strongly constraining \citep{roy_diversity_2025}, so an abundance of \ce{H2O} similar to that of \ce{CH4}, according to case 4, remains consistent with existing data. Therefore K2-18\,b, TOI-732\,c, TOI-270\,d and LP791-18\,c are all consistent with formation beyond the water ice line. 

In contrast to the rest of the population, GJ9827\,d has a high mean molecular weight atmosphere and \ce{H2O} detected at a volume mixing ratio $>$\,30\% \citep{piaulet-ghorayeb_jwstniriss_2024}. This cannot be explained by a formation beyond the water ice line as that would result in a high \ce{CH4} abundance (case 3) or an intermediate mean molecular weight (case 4), however it can naturally be explained by a formation within the ice line as case 1 or 2, followed by fractionated mass loss enriching the \ce{H2O}/\ce{H2} ratio \citep{valatsou_oxygen_2026}. This interpretation is supported by GJ9827\,d's proximity to the edge of the radius valley, and the observations of high energy stellar flaring activity during the transit observations \citep{piaulet-ghorayeb_window_2025}. The case of GJ9827\,d exemplifies how our results challenge the conventional assumption that a water-rich atmosphere indicates formation beyond the water ice line: a methane-free water-rich atmosphere is, counterintuitively, a signature of formation inside it.

For the sub-Neptunes that have featureless transmission spectra, it is more difficult to determine a classification because a featureless spectrum is degenerate between a high mean molecular weight atmosphere (case 3) and high-altitude aerosols. Despite their featureless spectra, some constraints have still been placed on the atmospheric mean molecular weights. For TOI-836\,c a mean molecular weight lower than $\sim$\,6\,g\,mol$^{-1}$ could be ruled out based on the amplitude of the spectral features, however if the spectrum is instead being truncated by high altitude aerosols, the mean molecular weight could be as low as 4.25 g\,mol$^{-1}$ \citep{wallack_jwst_2024}. TOI-836\,c is therefore unlikely to belong to cases 1 or 2, but can't yet be distinguished between cases 3 or 4. For GJ436\,b, and TOI-776\,b and c, observations of the \ce{He} escape rate have been used to infer upper limits on their atmospheric mean molecular weights of $<$\,10.2, $<$\,13.6, and $<$\,12.8\,g\,mol$^{-1}$ respectively \citep{rogers_using_2026}, tentatively ruling out case 3 for each planet, unless the atmospheric vertical mixing is low (Figure \ref{fig:profiles}). TOI-776\,c additionally has a lower limit on mean molecular weight of $>$\,6\,--\,8\,g\,mol$^{-1}$ if high altitude aerosols are not responsible for the featureless spectrum \citep{teske_jwst_2025}, which could suggest case 3 with a low vertical mixing strength when combined with the upper limit from He escape. GJ3090\,b has also been a subject of interest for breaking the cloud-mean molecular weight degeneracy \citep{ahrer_escaping_2025}, leading to a lower limit of $>$\,7.1\,g\,mol$^{-1}$, however if aerosols can be formed around $\mu$bar pressures then the mean molecular weight remains unconstrained \citep{parker_limits_2025}.

In general, distinguishing cases 1 and 2 (formation within the water ice line) from cases 3 and 4 (formation beyond the water ice line) can be done based on mean molecular weight, with the added consideration of \ce{H2O} condensation in the colder population, exemplified by K2-18\,b. Distinguishing case 1 from case 2 is more ambiguous and will require better observational and modeling constraints on the wider carbon chemistry involving \ce{CO} and \ce{CO2}, which is dependent not only on the different formation cases but also on the atmospheric temperature profile, vertical mixing strength, and photochemistry (Figure \ref{fig:profiles}). Distinguishing case 3 from case 4 might appear straightforward due to the different mean molecular weight atmospheres that they predict, however equilibrium chemistry at different T$_{\text{AMOI}}$, and disequilibrium chemistry over the vertical extent of the atmosphere can each significantly alter the mean molecular weight of case 3 atmospheres, blurring the observational identifiers of soot-water-rock formation with only water-rock formation scenarios (Figure \ref{fig:overview}). Unambiguously, however, \ce{CH4}-rich atmospheres can only be achieved by formation beyond the water ice line, while \ce{H2O}-rich atmospheres without significant \ce{CH4} are counterintuitively indicators of formation within the water ice line.

In summary, Figure~\ref{fig:overview} reveals a clear compositional trend from volatile-poor to volatile-rich formation scenarios. The equilibrium solution space traces an approximately linear sequence in log-log space, where  \ce{H2O}/\ce{CH4} and MMW serve as the primary diagnostic: low \ce{H2O}/\ce{CH4} and high MMW identify volatile-rich accretion beyond the water ice line (cases 3 and 4), while high \ce{H2O}/\ce{CH4} and low MMW indicate volatile-poor accretion within the water ice line (cases 1 and 2). Two processes can cause planets to deviate from this in-equilibrium trend. First, in sufficiently cool atmospheres, \ce{H2O} condensation removes water from the observable upper atmosphere, shifting apparent \ce{H2O}/\ce{CH4} ratios to lower values, as observed for K2-18\,b. Second, fractionated atmospheric escape preferentially removes hydrogen, elevating the mean molecular weight and displacing planets toward higher MMW at fixed \ce{H2O}/\ce{CH4}, as possibly exemplified by GJ9827\,d near the radius valley. Accounting for these two effects, the \ce{H2O}/\ce{CH4} ratio and mean molecular weight together provide a powerful two-dimensional diagnostic for sub-Neptune formation history that is broadly consistent with the current JWST population.

\subsubsection{Carbon dioxide as additional discriminant in K2-18\,b and TOI-270\,d}

The two well-characterized sub-Neptunes K2-18\,b and TOI-270\,d provide constraints on
carbon- and oxygen-bearing species beyond \ce{H2O} and \ce{CH4}, and in particular on
\ce{CO2}, which our models suggest may help discriminate between formation scenarios.

For K2-18\,b, multiple independent retrieval studies consistently report significant
\ce{CO2} detections, with inferred log-abundances spanning roughly $-6$ to $-2$ and C/O
ratios ranging from moderately to strongly super-solar
\citep{fernandez-rodriguez_atmospheric_2025, schmidt_comprehensive_2025}.
For TOI-270\,d, the retrieved carbon-species abundances are more tightly clustered, with
both log(\ce{CO2}) and log(\ce{CH4}) typically falling between $-3$ and $-1$.
C/O ratios are generally approximately solar to super-solar, with individual retrieval
cases reaching values as high as $3.21^{+6.1}_{-1.9}$ \citep{felix_evidence_2025}, and
atmospheric metal mass fractions are consistently high, with representative values around
$Z_\mathrm{atm} \sim 0.6$.

The \ce{CO2} abundances further confirm the conclusion drawn from \ce{H2O}/\ce{CH4}
in Section~\ref{sec:Discussion}: both planets strongly disfavour water-poor formation
pathways (cases~1 and~2).
In water-poor scenarios, carbon is sequestered deep in the planetary interior rather than
outgassed into the atmosphere, keeping log(\ce{CO2}) $\lesssim -6$ and placing it well
outside the observational constraints for both planets.
Only water-rich initial compositions reproduce the observed \ce{CO2} abundances, C/O
ratios, and atmospheric metal mass fractions within the correct order of magnitude.
We note that this conclusion differs from \citet{nixon_magma_2025}, who employed a similar interior–atmosphere chemical equilibrium framework but without carbon partitioning into the metallic phase. In that end-member treatment, carbon was mainly present by construction in the atmosphere, which naturally promotes higher atmospheric carbon abundances in volatile-poor scenarios. Our treatment explicitly allows carbon to partition into the metal, revealing that sequestration into the deep interior becomes the dominant fate of carbon when volatile-poor building blocks are assumed. Both approaches confirm that magma ocean–atmosphere equilibration can reproduce the observed atmospheric composition of planets such as TOI-270\,d; the two frameworks differ primarily in the predicted carbon budget of volatile-poor formation scenarios, where the inclusion of metal–carbon partitioning shifts carbon away from the atmosphere and into the interior.

\ce{CO2} also offers a potential discriminant between cases~3 and~4.
Case~4 predicts log(\ce{CO2}) between roughly $-5$ and $-2$ at low atmospheric pressures
(Figure~\ref{fig:profiles}), whereas case~3 tends to yield lower values spanning
approximately $-8$ to $-3$.
For TOI-270\,d, the relatively high and tightly constrained \ce{CO2} abundance, together
with sub-unity C/O ratios, appears more consistent with case~4, suggesting that a soot
component is not required to explain its atmosphere.
For K2-18\,b, the picture is less clear: while some retrievals are compatible with the
higher C/O and lower \ce{CO2} characteristic of case~3, the low mean molecular weight
favours case~4.
This preference is reinforced by interior modeling suggesting a possible magma ocean interface
temperature of $T_\mathrm{AMOI} \approx 3000\,$K or higher for K2-18\,b \citep{rigby_toward_2024}; at this
temperature, case~4 produces a tight MMW distribution consistent with observations,
whereas reducing $T_\mathrm{AMOI}$ toward $\sim$2000\,K introduces a large spread in
predicted MMW (Figure~\ref{fig:overview}) that is harder to reconcile with observational constraints.

Taken together, the available evidence tentatively suggests that both K2-18\,b and
TOI-270\,d can be explained without invoking a significant soot component.
However, given the substantial observational uncertainties and model dependencies in
current retrievals, this interpretation remains speculative.
A tighter constraint on $T_\mathrm{AMOI}$ from interior modeling, and improved retrieval
precision on \ce{CO2} and C/O for K2-18\,b in particular, would help to more robustly
distinguish cases~3 and~4.

\section{Caveats and Limitations}
\label{limit}

The results presented here rest on two main idealizing assumptions.

The most fundamental is complete interior–atmosphere chemical equilibration. Convective inhibition and mean molecular weight gradients may restrict chemical exchange on a global scale \citep[e.g.,][]{leconte_condensation-inhibited_2017,spaargaren_influence_2020,misener_importance_2022,2022Markham,2024Leconte3d}. In particular, mean molecular weight gradients in the atmosphere may operate over long timescales and limit how far one can make interpretations about the bulk composition from the observable part of the atmosphere. The true atmospheric composition of any individual planet likely lies between the fully equilibrated state modeled here and the chemically decoupled interior assumed in previous soot-planet models. The vertical mixing sensitivity explored in Figure \ref{fig:profiles} only partially accounts for this limitation; ideally, evolutionary models that track compositional gradients, their erosion, and layered convection \citep{2022Markham} would be employed, but this is beyond the scope of the present work.

A second key uncertainty is the magma ocean interface temperature $T_\mathrm{AMOI}$. We adopt 3000 K as our fiducial value and demonstrate that the main compositional trends are preserved at 2000 K, though the predicted MMW distribution for case 4 significantly broadens at lower temperatures, reducing the discriminating power of that diagnostic. The choice of upper bound is physically motivated: recent work suggests that at temperatures approaching 4000 K, gases and refractories become miscible \citep{young_differentiation_2025}, dissolving the distinct magma ocean–atmosphere interface that our model requires. The thermodynamic data underlying our GCE model would also require substantial extrapolation beyond this temperature. At the lower bound, a magma ocean ceases to exist below approximately 2000 K. We therefore consider our tested cases with 2000 and 3000 K to be a physically reasonable and representative temperature range for the AMOI conditions of sub-Neptunes.

\section{Conclusions} \label{sec:Conclusions}

We have shown that the atmospheric compositions of sub-Neptunes carry a direct imprint of their formation environment, and that global chemical equilibration between planetary interiors and atmospheres is essential for correctly interpreting this imprint. By modeling four end-member core compositions — rock, soot--rock, soot--water--rock, and water--rock — spanning the range of building blocks available at different locations in the protoplanetary disk, we identify a set of atmospheric diagnostics that link JWST observables to planetary origins.

Planets formed from water-poor material produce atmospheres strongly depleted in carbon-bearing species, with log(\ce{CH4}) and log(\ce{CO2}) below -5, consistent with TOI-421\,b and GJ3470\,b, while GJ9827\,d may have evolved from such a state through preferential hydrogen loss. Planets assembled from water-rich building blocks naturally yield \ce{CH4} and \ce{CO2} rich atmospheres with elevated metal mass fractions and C/O ratios, consistent with TOI-776\,c, TOI-270\,d, TOI-836\,c, and LP791-18\,c. In water-poor scenarios, carbon is sequestered deep in the planetary interior rather than outgassed, a result that depends critically on allowing carbon to partition into the metallic phase -- a key difference from otherwise similar magma ocean frameworks \citep{nixon_magma_2025}-- and explains why only water-rich formation reproduces the observed \ce{CO2} abundances, C/O ratios, and atmospheric metal mass fractions of TOI-270\,d as well as K2-18\,b within the correct order of magnitude.

\ce{CO2} further serves as a useful discriminant between soot-bearing (case\,3) and soot-free water-rich (case\,4) formation: case\,4 predicts log(\ce{CO2}) between roughly -5 and -2, whereas case~3 tends to yield lower values. The tightly constrained \ce{CO2} abundance and sub-unity C/O ratios of TOI-270\,d appear more consistent with case\,4, as does the low mean molecular weight of K2-18\,b. However, the latter conclusion remains speculative given current observational uncertainties and its dependence on the assumed magma ocean interface temperature. Soot-bearing interiors can further enhance methane abundances and produce methane-dominated atmospheres, as possibly inferred for TOI-836\,c if aerosol explanations can be ruled out. Additionally, the water-rich case\,4 may help explain the long-standing non-detection of \ce{CO} in sub-Neptune atmospheres: although \ce{CH4}, \ce{H2O}, and \ce{CO} are dominant at depth, \ce{CO} abundances can decrease by up to eight orders of magnitude under weak vertical mixing, compared to at most five orders of magnitude for \ce{CO2}, suggesting that retrieved compositions carry information on both primordial building blocks and mixing efficiency.

Together, these case-by-case comparisons motivate a more general diagnostic. We demonstrate how the \ce{H2O}/\ce{CH4} versus MMW parameter space provides a powerful and physically motivated diagnostic linking atmospheric observables to planetary building blocks, with the four formation cases occupying distinct regions separable at the order-of-magnitude level. Outliers from the predicted trends are naturally explained by fractionated atmospheric escape, as inferred for GJ9827\,d \citep{piaulet-ghorayeb_jwstniriss_2024, krishnamurthy_absence_2023, valatsou_oxygen_2026}, and water condensation, as inferred for K2-18\,b \citep{hu_water-rich_2025}. Interestingly, representative retrieved atmospheres exist for almost all modeled cases, suggesting that sub-Neptunes may form across a broader range of disk environments than theoretical arguments favouring formation beyond the water ice line would imply \citep{venturini_nature_2020}.

Looking forward, the most decisive observational test is a precise mean molecular weight measurement for a planet currently classified as case 3 (with soot) or case 4 (without soot) candidate. Case 3 and 4 predict MMW $>$17 and $<$7.5 respectively, which is a factor of two difference that is in principle measurable from the amplitude of spectral features across JWST's full wavelength range, or from atmospheric escape observations sensitive to He and H. On the modeling side, the key uncertainty is the magma ocean interface temperature $T_\mathrm{AMOI}$ which controls the predicted MMW spread for case 4. Interior structure models that jointly fit mass, radius, and atmospheric composition would help narrow this parameter and sharpen the predicted observational boundaries between cases. For planets where \ce{CO2} is already detected, such as K2-18\,b, improved retrieval precision on the \ce{CO2} abundance and C/O ratio, which is feasible with additional JWST transits, would help resolve whether a soot component is required or whether water-rich formation alone suffices. Finally, the \ce{H2O}/\ce{CH4} vs MMW diagnostic framework presented here should be applied systematically to the growing JWST sub-Neptune sample as new characterizations become available.
The diversity of sub-Neptune atmospheres is ultimately a record of their formation history, but it can only be decoded by explicitly accounting for the chemical equilibration between planetary interiors and their atmospheres.

\section*{Acknowledgements}

C.D acknowledges support from the Swiss National Science Foundation under grant TMSGI2\_211313. This work has been carried out within the framework of the NCCR PlanetS supported by the Swiss National Science Foundation under grant 51NF40\_205606. We acknowledge the use of large language models (LLMs), including ChatGPT, to improve the grammar, clarity, and readability of the manuscript. S.J. acknowledges funding support from ETH Zurich and the NOMIS Foundation in the form of a research fellowship. The NOMIS Foundation ETH Fellowship Programme and respective research are made possible thanks to the support of the NOMIS Foundation.

\section*{ORCID iDs}

\noindent 
Caroline Dorn \orcidlink{0000-0001-6110-4610} \href{https://orcid.org/0000-0001-6110-4610}{0000-0001-6110-4610} \\
Aaron Werlen \orcidlink{0009-0005-1133-7586} \href{https://orcid.org/0009-0005-1133-7586}{0009-0005-1133-7586} \\
Sean Jordan \orcidlink{0000-0002-2828-0396} \href{https://orcid.org/0000-0002-2828-0396}{0000-0002-2828-0396} \\

\begin{appendix}
 \nolinenumbers
\section{thermodynamic network}
\label{sec:appendix_equilibrium}

The GCE model from \citep{grimm_new_2026} is adapted from the original model in \cite{schlichting_chemical_2022} with the addition of carbon following \cite{werlen_atmospheric_2025, werlen_sub-neptunes_2025}. In total, the chemical network is defined with 19 linearly independent reactions involving 26 phase components. Chemical reactions are allowed within the gas and silicate phases as well as between the metal-silicate-gas phases.

Reactions within the silicate phase are:

\begin{equation}
    \ce{Na2SiO3_{,silicate} \rightleftharpoons Na2O_{silicate} + SiO2_{,silicate}} \tag{R1}
\end{equation}

\begin{equation}
    \ce{MgSiO3_{,silicate} \rightleftharpoons MgO_{silicate} + SiO2_{,silicate}} \tag{R2}
\end{equation}

\begin{equation}
    \ce{FeO_{silicate} + 1/2Si_{metal} \rightleftharpoons Fe_{metal} + 1/2SiO2_{,silicate}} \tag{R3}
\end{equation}

\begin{equation}
    \ce{FeSiO3_{,silicate} \rightleftharpoons FeO_{silicate} + SiO2_{,silicate}} \tag{R4}
\end{equation}

Reactions between the metal and the silicate phase include:

\begin{equation}
    \ce{O_{metal} + 1/2Si_{metal}} \rightleftharpoons \ce{1/2SiO2_{,silicate}} \tag{R5}
\end{equation}

\begin{equation}
    \ce{2H_{metal} \rightleftharpoons H2_{,silicate}} \tag{R6}
\end{equation}

\begin{equation}
    \ce{Si_{metal} + 2H2O_{silicate} \rightleftharpoons SiO2_{,silicate} + 2H2_{,silicate}} \tag{R7}
\end{equation}

\begin{equation}\label{Carbonreaction}
    \ce{C_{metal} + O_{metal} \rightleftharpoons CO_{silicate}} \tag{R8}
\end{equation}

Reactions among gas-phase species are:

\begin{equation}
    \ce{CO_{gas} + 1/2O2_{,gas} \rightleftharpoons CO2_{,gas}} \tag{R9}
\end{equation}

\begin{equation}
    \ce{CH4_{,gas} + 1/2O2_{,gas}} \rightleftharpoons \ce{2H2_{,gas} + CO_{gas}} \tag{R10}
\end{equation}

\begin{equation}
    \ce{H2_{gas} + 1/2O2_{,gas} \rightleftharpoons H2O_{gas}} \tag{R11}
\end{equation}

Reactions representing magma ocean–atmosphere exchange are:

\begin{equation}
    \ce{FeO_{silicate} \rightleftharpoons Fe_{gas} + 1/2O2_{,gas}} \tag{R12}
\end{equation}

\begin{equation}
    \ce{MgO_{silicate} \rightleftharpoons Mg_{gas} + 1/2O2_{,gas}} \tag{R13}
\end{equation}

\begin{equation}
    \ce{SiO2_{,silicate} \rightleftharpoons SiO_{gas} + 1/2O2_{,gas}} \tag{R14}
\end{equation}

\begin{equation}
    \ce{Na2O_{silicate}} \rightleftharpoons \ce{2Na_{gas} + 1/2O2_{,gas}} \tag{R15}
\end{equation}

\begin{equation}
    \ce{H2_{,silicate} \rightleftharpoons H2_{,gas}} \tag{R16}
\end{equation}

\begin{equation}
    \ce{H2O_{silicate} \rightleftharpoons H2O_{gas}} \tag{R17}
\end{equation}

\begin{equation}
    \ce{CO_{silicate} \rightleftharpoons CO_{gas}} \tag{R18}
\end{equation}

\begin{equation}
    \ce{CO2_{,silicate} \rightleftharpoons CO2_{,gas}} \tag{R19}
\end{equation}

Note that any additional reaction that can be formed as a linear combination of these basis reactions is also permitted. The reaction space is therefore not limited to the expressions explicitly listed above.

Chemical equilibrium among the 19 reactions is determined by solving the following condition for the molar fractions $x_i$ of all participating species:

\begin{equation}\label{eq:chemical_equilibrium}
    \sum_i \nu_i \ln x_i + \left[\frac{\Delta \hat{G}^\circ_{\text{rxn}}}{RT} + \sum_g \nu_g \ln(P/P^\circ)\right] = 0,
\end{equation}

\noindent where $x_i$ denotes the mole fraction of species $i$ within its respective phase, and $\nu_i$ are the associated stoichiometric coefficients. The term $\Delta \hat{G}^\circ_\text{rxn}$ represents the standard Gibbs free energy change of the reaction, $R$ is the ideal gas constant, $T$ is the temperature, $P$ is the pressure at the atmosphere–magma ocean interface (AMOI), and $P^\circ$ is the reference pressure, set here to 1 bar. The summation over $g$ accounts for gas-phase species only, which introduces the explicit pressure dependence.

In addition, the system includes seven elemental mass balance constraints for H, C, O, Na, Si, Mg, and Fe, as well as three phase normalization conditions for the gas, silicate, and metal phases. The complete system therefore comprises 29 coupled nonlinear equations. These equations are solved simultaneously for the mole fractions $x_{i,k}$ of the 26 species and the total mole numbers $N_k$ of the three phases.

The thermodynamic parameters underlying the chemical reaction network are adopted from the compilation presented in \citet{schlichting_chemical_2022}. In addition to their framework, we explicitly account for carbon dissolution into the metallic phase by implementing metal–silicate partition coefficients from \citet{blanchard_metal-silicate_2022}. Hydrogen solubility in silicate melt is described using the revised formulation introduced by \citet{werlen_effects_2026}, which combines the experimental constraints of \citet{hirschmann_magma_2012} with DFT–MD results from \citet{gilmore_coreenvelope_2026}.

Deviations from ideal solution behavior are considered only within the metallic phase. For Si, O, and H, we follow the interaction model outlined in \citet{young_earth_2023} and applied in \citet{werlen_sub-neptunes_2025}, ultimately based on the formalism of \citet{badro_core_2015}. Carbon activity coefficients in metal are taken from \citet{fischer_carbon_2020}, consistent with the implementation described by \citet{werlen_sub-neptunes_2025}. We assume ideal mixing in both the gas and silicate melt phases. As discussed by \citet{werlen_effects_2026}, introducing non-ideality in only a subset of phases can produce artificial compositional trends, while robust high-pressure, high-temperature activity data for multicomponent silicate melts remain limited.

\end{appendix}
\newpage
\bibliography{references}
\bibliographystyle{aasjournal}

\end{document}